\documentstyle[prl,aps,epsf,cite,multicol,graphicx]{revtex}

\newcommand{\be}{\begin{equation}}
\newcommand{\ee}{\end{equation}}
\newcommand{\bea}{\begin{eqnarray}}
\newcommand{\eea}{\end{eqnarray}}
\newcommand{\nn}{\nonumber}

\newcommand{\rr}{{\bf r}}
\newcommand{\qq}{{\bf q}}
\newcommand{\kk}{{\bf k}}

\newcommand{\rmd}{{\rm d}}

\begin{document}
 
\flushbottom

\draft


\title{Liquid crystalline states for two-dimensional electrons in
    strong magnetic fields}

\author{Orion Ciftja}
\address{Department of Physics, Prairie View A\&M University,
         Prairie View, Texas 77446, USA}

\author{Cintia M.\ Lapilli}
\address{Department of Physics and Astronomy,
         University of Missouri--Columbia,
         Columbia, Missouri 65211, USA}

\author{Carlos Wexler}
\address{Department of Physics and Astronomy,
         University of Missouri--Columbia,
         Columbia, Missouri 65211, USA}

\date{October 13, 2003}

\maketitle


\begin{abstract}
Based on the Kosterlitz-Thouless-Halperin-Nelson-Young (KTHNY) theory of
two-dimensional melting and the analogy between Laughlin states
and the two-dimensional one-component plasma (2DOCP), we investigate
the possibility of liquid crystalline states in a single Landau
level (LL).
We introduce many-body trial wavefunctions that are translationally
invariant but posess 2-fold (i.e. {\em nematic} ), 4-fold ({\em
tetratic}) or 6-fold ({\em hexatic}) broken rotational symmetry at
respective filling factors $\nu = 1/3$, 1/5 and 1/7 of the valence LL.
We find that the above liquid crystalline states exhibit a soft charge
density wave (CDW) which underlies the translationally invariant state but
which is destroyed by quantum fluctuations.
By means of Monte Carlo (MC) simulations, we determine that, for a
considerable variety of interaction potentials, the anisotropic states
are energetically unfavorable for the lowest and first excited LL's
(with index $L = 0, 1$), whereas the nematic is favorable at the second
excited LL ($L = 2$).


\end{abstract}
%

\begin{multicols}{2}
\section{Introduction}
\label{sec:introduction}

\noindent
In 1983 Laughlin \cite{laughlin83} introduced his famous trial wave
function 
\be
\label{eq:lauglin}
\Psi_{1/m}= \prod_{i<j}^{N} \, (z_i-z_j)^{m} \,
         e^{-\frac{1}{4 l_0^2}\sum_{k=1}^{N} |z_k|^2} \,,
\ee
to describe the fractional quantum Hall effect(FQHE)
states \cite{tsui82,perspectives96,heinonen,jain} for filling factors
$\nu=1/m$ of the lowest Landau level (LLL), where $m$ is an odd
integer.  Immediately after this discovery, many attempts were done to
compare the stability of these states against other known ground
states, typically  Wigner crystal (WC) states \cite{mendez83,liqusolid,wig2}.
At absolute zero ($T = 0$) the current theoretical understanding is
that WC states are favorable for filling factors smaller than a
critical value $\nu_c \simeq {1}/{6.5}$ \cite{liqusolid,wig2}.  For
larger filling factors of the LLL, the electrons are believed to form
a quantum liquid state with Laughlin wave function being an excellent
choice for $\nu=1/m$ (with $m = 1,3,5$) \cite{iqhe}.
Because of its translational and rotational invariance,
Laughlin's wave function can be used to describe a liquid
state of the electrons in the LLL, as can be seen
by writing  $ |\Psi_{1/m}|^2 $ as a classical distribution function
\cite{caillol,levesque}:
\bea
\label{2docp}
&&|\Psi_{1/m}|^2 \propto e^{-\beta V}  \ , \ {\rm where}\\
&&-\beta V = 2m \sum_{i<j}^{N} \, \ln |z_i-z_j|-\frac{1}{2 l_0^2}
             \sum_{k=1}^{N} |z_k|^2  \ , \nn
\eea
and $V$ is the potential energy of a classical two-dimensional
one-component-plasma (2DOCP) system. Using the formal analogy between
the Laughlin wave function and the 2DOCP we can identify a
dimensionless coupling constant, $\Gamma \equiv \beta e^2 = e^2/(k_B T) = 2m$.
An equlibrium state of the 2DOCP is
entirely characterized by $\Gamma$ and the freezing transition in this
case was located at $\Gamma \approx 140$ \cite{caillol}.
Employing the analogy between the temperature of the classical plasma
and the  filling factor of the LLL, we should expect a freezing
transition as we decrease the electronic filling factor in the quantum
Hall regime. Because of the different quantum nature of the electronic
correlations in the FQHE, it was found that such a system is a
Laughlin liquid for filling factors $\nu=1/3$ and $1/5$, but becomes
crystal for filling factors smaller than $\nu_c \approx 1/6.5$ (this
value is about an order of magnitude larger than that deduced from
 the classical 2DOCP analogy).

It is feasible that, in analogy to the classical freezing transition
realized by cooling down a 2DOCP, the transition to a solid (WC) state
obtained by reducing the filling factor in the electron case may be
interpreted as a topological Kosterlitz-Thouless-type transition
\cite{kt}.  This would be the correlated electron system
counterpart of the wellknown 2D melting problem. Although the 2D
melting is not fully understood, an elegant and reliable theory of
melting has been proposed in the 1970's by Kosterlitz, Thouless,
Halperin, Nelson and Young (KTHNY) \cite{kt,hn,young}. The KTHNY
theory predicts that an intermediate {\em third phase} called
{\em hexatic}, will exist between the hexagonal solid and the liquid phases
in a certain portion of the phase diagram (perhaps in a somewhat
narrow range of temperatures).  In the liquid phase there is no
long-range translational or rotational order (the system is both
translationally and rotationally invariant). In the solid phase the
system has quasi-long-range translational and true long-range
rotational order. The hexatic phase in the KTHNY theory is thought to
have no true long-range translational order, but does retain
quasi-long-range orientational order (the system is translationally
invariant, but not rotationally invariant at least for short
distances).  The intermediate hexatic phase is often considered most
important since it has a symmetry intermediate between the hexagonal
solid and the liquid.

Recent experiments in very high mobility ($\mu \sim 10^7$ m/Vs)
{GaAs/Al$_x$Ga$_{1-x}$As} heterostructures have shown a variety of low
temperature phases with exotic properties.  Since 1999 it has been
known that in transitional regions between QH plateaus for high
LL's (with LL index $L \ge 2$) either a {\em smectic} or {\em nematic}
phase exists
\cite{anisotropic,fradkin,wexlerKT,brsthird,brshalf,cw2002}.  In fact,
one of us calculated to a reasonable accuracy the anisotropic-isotropic
transition temperature as a topological process \cite{wexlerKT}.  In
2002 a melting transition from the WC state to a FQHE-like state was
observed at ca.\ 130 mK \cite{melting17} and speculation mounted to
suggest that possibly this transition occurs to a hexatic mesophase
\cite{hexatic}.

On this grounds we investigate the possibility of various liquid
crystalline mesophases in a partially filled LL.  Given that
two-dimensional liquid crystals may posses different forms of
rotational group symmetry, we select a set of possible
candidates, having $C_2$ (nematic), $C_4$ (tetratic), and $C_6$
(hexatic) rotational group symmetry (note that in principle higher
symmetry groups are also possible for a liquid crystal, e.g.\ a {\em
liquid quasicrystal} with a $C_{10}$ symmetry---we have not explored,
however such possibilities in this paper \cite{lqc}).  Our results indicate
that the states studied exhibit a soft charge density wave (CDW) which
underlies the translationally invariant state but which is
destroyed by quantum fluctuations.  We perform Monte Carlo (MC) simulations
and determine that, for a wide range of interactions the anisotropic states
are energetically unfavorable for the lowest and first excited LL's
(with index $L = 0, 1$), whereas the nematic is favorable at the second
excited LL ($L = 2$).

In Sec.\ \ref{sec:trial_states} we describe the types of states that
were considered for our calculations.  Section \ref{sec:MC} presents the
types of interaction potential considered and explains the methods
used to calculate the properties of the system.  Section
\ref{sec:results} contains the results obtained and a discussion of
their meaning.  The underlying soft CDW is discussed in Sec.\ \ref{sec:cdw}.
Finally, the conclusions are presented in Sec.\ \ref{sec:conclusions}.

\section{Liquid crystal states}
\label{sec:trial_states}

\noindent
In this paper we consider liquid crystalline phases with no
translational order but with quasi-long-range orientational order with
various rotational symmetry groups $C_2$, $C_4$, and $C_6$;
corresponding to a {\em nematic}, {\em tetratic} and {\em hexatic}
phase respectively.
There are some basic requirements on how we construct these states:
  {\em (i)}  the states must obey Fermi
statistics, i.e.\ they must have odd parity under the exchange of any pair of
electrons; {\em (ii)} the states must be translationally invariant (at
least far away from the boundaries of the system in case of a finite
number of electrons); {\em (iii)} there must be a broken rotational
symmetry belonging to the proper symmetry group; {\em (iv)} the states
must belong to a single LL to avoid the large cyclotron energy
cost $\hbar \omega_c = \hbar e B/m_e$, where $B$ is the magnetic field,
and $e$ and $m_e$ are the electron charge and mass respectively (also
note that as we will show later, various properties at {\em any} LL
can be readily obtained from properties calculated in the LLL).

A class of such wave functions satisfying all these requirements are the
so-called broken-rotational-symmetry (BRS) wave
functions \cite{joynt,brsthird,brshalf,cw2002,hexatic} that are
systematically constructed by properly splitting the zeros of the
Laughlin liquid state [in essence, the idea is to place the vortices that
perform the composite fermion (CF) transformation \cite{heinonen,jain}
{\em around} the location of the electron, rather ``on top'' of them].
Let us consider the Laughlin wavefunction as
given in Eq.\ (\ref{eq:lauglin}),
where $z_k = x_k + i y_k$ is $k$-th electron position in
the $xy$-plane in complex notation, and $l_0=[\hbar/(e B)]^{1/2}$ is
the magnetic length.  This wave function represents a gaped, uniform and
isotropic liquid, and is an excellent description of a liquid state at
filling factor $\nu = 1$, 1/3 and 1/5 of the LLL (for $\nu = 1/7$, the
WC state prevails, see previous discussion, and Ref.\
\onlinecite{hexatic}).

To build a liquid crystal (BRS) state out of the liquid states
we split the zeros of the wave function in a way that conserves
the anti-symmetry (Fermi statistics) and translational invariance,
but breaks the rotatational invariance of the wave function.
This is done by introducing a prefered set of directions
\cite{joynt,brsthird,brshalf,cw2002,hexatic} into the wave function
creating a degree of anisotropy. A generalized liquid crystal wave
function for a filling factor $\nu=1/m$ can then be easily written as:
\bea
\label{eq:psi_nematic}
\Psi_{1/m}^\alpha &=&
         \left\{ \prod_{i<j}^{N}
        \left[ \prod_{\mu=1}^{m-1} (z_i-z_j-\alpha_\mu) \right]
    \right\} \\ & &
    \times
     \prod_{i<j}^{N} \, (z_i-z_j) \,
        e^{-\frac{1}{4 l_0^2}\sum_{k=1}^{N} |z_k|^2} \,, \nn
\eea
where the complex directors $\alpha_\mu$ are distributed in pairs of
opposite value in the complex plane (to satisfy Fermi statistics).  In
this paper we focus on the states with the highest level of discrete
symmetry possible at each filling factor, which is set by distributing
the $\alpha_\mu$ symmetrically in a circle around the origin:
\be
    \alpha_\mu = \alpha \ e^{i \, 2\pi (\mu-1)/(m-1)} \,, \ \
    \mu \in \{1,2,\ldots, (m-1)\} \,.
\ee
Without loss of generality $\alpha$ can be taken to be real.
The wavefunction in Eq.\ (\ref{eq:psi_nematic})
 represents a homogeneous liquid crystalline state at
filling factors $\nu=1/m$, is anti-symmetric, lies entirely in the
LLL, and is smoothly connected to the isotropic Laughlin state for
$\alpha = 0$.

\section{Interaction potentials and Monte Carlo simulation}
\label{sec:MC}

For our simulations we consider $N$ electrons in a charge neutralizing
background.  When considering the quantum Hamiltonian
$\hat{H}=\hat{K}+\hat{V}$, the strong magnetic field quantizes the
kinetic energy $\hat{K}$ so that single-LL wavefunctions have a
constant (and thus irrelevant) kinetic energy, $\langle \hat{K} \rangle
/ N$.
The only relevant contribution comes, therefore, from the total
potential energy operator
\begin{equation}
\hat{V}=\hat{V}_{ee}+\hat{V}_{eb}+\hat{V}_{bb} \ ,
\label{pot_en}
\end{equation}
consisting of electron-electron, electron-background and
background-background interactions.

It has been a common practice to work on the surface of a sphere
\cite{jain} in order to minimize boundary effects in the finite-size
computations.  However, due to the anisotropic nature of the states
under consideration this scheme would produce significant problems due
to the need to have topological defects at the ``poles'' of the
sphere.  We therefore work on a simpler disk geometry, where the
neutralizing positive background has a uniform density $\rho_0=\nu/(2
\pi l_0^2)$ and is spread over a disk of radius $R_N = l_0 (2 N/\nu)^{1/2}$
with an area $\Omega_N = \pi R_N^2$.

%
Our goal is to thoroughly investigate the possibility of a liquid
crystal state in the LLL for electrons interacting not only with the
usual bare Coulomb potential $v_C(r_{12})=e^2/(\epsilon r_{12})$
but also for a variety of other reasonable effective potentials that
take into consideration the finite thickness of the quasi-2D electron
layer.  As previously shown by Zhang and Das Sarma (ZDS)\cite{ZDS},
the electron-electron interaction in a quasi-2D system can be written
as:
\bea
\label{thickness}
v_{ZDS}(r_{12}) &=& \frac{e^2}{\epsilon}
\int_{0}^{\infty} dq \, J_0(q \, r_{12}) \, F(q,b)   \ ,
   \nonumber    \\
  F(q,b) &=& \left(1+\frac{9}{8} \frac{q}{b}+\frac{3}{8}
\frac{q^2}{b^2} \right)
         \left(1+\frac{q}{b} \right)^{-3}    \ ,
\eea
where $r_{12}$ is the 2D distance separating the two electrons,
$\epsilon$ is the average background dielectric constant,
$J_0$ is the Bessel function of zeroth order, and $b$ is a parameter
related to the finite thickness of the 2D layer (if we define the
average thickness as $\overline{Z}$, then $b=3/\overline{Z}$).
In addition, we also consider two other interaction potentials:
\bea
\label{2other}
v_1(r_{12}) &=&\frac{e^2}{\epsilon} \ \frac{1}{\sqrt{r_{12}^2+\lambda^2}}
            \ , \nn \\
v_2(r_{12})&=&\frac{e^2}{\epsilon}  \
            \frac{1-\exp(-\frac{r_{12}}{\lambda})}{r_{12}} \ .
\eea
The two model potentials include the thickness effect
phenomenologically \cite{ZDS} through the length parameter,
$\lambda=\overline{Z}/2=1.5/b$. All the above potentials have the same
Coulomb behavior for large $r_{12}$, but differ from the bare Coulomb
potential for small $r_{12}$.

%
%

To consider the zero-temperature stability of the liquid crystal
states of Eq.\ (\ref{eq:psi_nematic}) with respect to the uniform
isotropic  liquid state counterparts, we performed extensive MC
simulations in order to compute the energy and other
quantities for the four different interaction potentials.  Since the
potentials involved are merely single- and two-body interactions, we
need to accurately determine all single- and double-particle
distribution functions, i.e.\ the {\em density}
$\rho(\rr) \equiv \left\langle \sum_{i=1}^{N}
        \delta(\rr_i-\rr) \right\rangle $,
and the {\em pair correlation function} $g(\rr_{12})$, respectively.
The determination of such functions allows an accurate determination
of all potential energies  in the $N \rightarrow \infty$ thermodynamic
limit \cite{mc_methods}.

%
By definition, the pair correlation function, $g(\rr_{12})$ is the
conditional probability [normalized so that $g(\infty) = 1$]
to find an electron at position $\rr''$ given
that another electron is found at position $\rr' = \rr''-\rr_{12}$:
\be
\label{eq:gr}
    g(\rr_{12}) \equiv \frac{1}{\rho_0^2}
        \left \langle \sum_{i \neq j}^{N}
        \delta(\rr_i-\rr')\delta(\rr_j-\rr'') \right \rangle \,,
\ee
where $\rho_0=\nu/(2 \pi l_0^2)$ is the average bulk electron density.
It is also useful to define the static structure factor $S(\qq)$,
which is given by the 2D Fourier transform of
$g(\rr_{12})$:
\be
\label{eq:sq}
S(\qq) - 1 =  \rho_0
        \int d^2r_{12} \, e^{-i \qq \cdot \rr_{12}} \, [g(\rr_{12}) - 1] \,.
\ee
Note that, because of the anisotropy of the wave function, both
functions are explicitly angle-dependent:
$g(\rr_{12})=g(r_{12},\theta)$ and $S(\qq)=S(q,\theta_q)$ for $\alpha
\neq 0$.  It is also worth noting that the charge neutrality sum rule
guarantees that $S(\qq) \propto q^2$ for $q\rightarrow 0$
\cite{caillol,SMA,sumrules}.

In the thermodynamic limit, the ground state correlation energy per
particle can be easily computed from \cite{mc_methods}:
\be
\label{eq:ener}
E_{\alpha}= \frac{\langle \hat{V}\rangle}{N} =
    \frac{\rho_0}{2} \int d^2r_{12} \, v(r_{12}) \,
       \left[ g(\rr_{12})-1 \right] \ ,
\ee
where $v(r_{12})$ can have any reasonable form, in particular it can
take the form of any of the potentials shown in Eqs.\
(\ref{thickness},\ref{2other}).
Because the interaction potentials are centrally symmetric,
the above formula can be rewritten in the simpler form:
\be
\label{eq:energy}
E_{\alpha}= 
    \frac{\rho_0}{2} (2 \pi)  \int_{0}^{\infty} dr_{12} \, r_{12} \,
      v(r_{12}) \,
       \left[ \overline{g}(r_{12})-1 \right] \ ,
\ee
where
$\overline{g}(r_{12})$ is the angle-averaged pair distribution function:
\be
\label{angleav-g}
\overline{g}(r_{12}) = \int_{0}^{2 \pi} \frac{\rmd \theta}{2 \pi} \;
g(\rr_{12}) \ .
\ee

For specific cases [as for the $v_{ZDS}(r_{12})$ potential,
which has strongly oscillatory behavior in real space, making the
numerical calculations very unstable and precarious] a corresponding
formula that uses the static structure factor was employed:
\be
\label{eq:energy-struc}
E_{\alpha}= 
    \frac{1}{2} \frac{1}{(2 \pi)}  \int_{0}^{\infty} dq \, q \,
      \tilde{v}(q) \,
       \left[ \overline{S}(q)-1 \right] \ .
\ee
In this case $\tilde{v}(q)$ is the 2D Fourier transform of the interaction
potential, and we also define the angle-averaged static structure factor
$\overline{S}(q) =  \int_{0}^{2 \pi} {\rmd \theta_{q}}/({2
\pi}) \; S(\qq)$.  The use of the static structure factor has the
added advantage of allowing the calcuation of the correlation energies
in {\em all } LL's from a single determination of the
pair correlation function in the LLL \cite{brsthird,brshalf,cw2002}:
\be
\label{eq:energy-struc-L}
E_{\alpha}^{(L)}=
    \frac{1}{2}   \int_{0}^{\infty} \!\! \frac{dq}{(2 \pi)} \, q \,
      \tilde{v}(q) \, [L_L(\frac{q^2}{2})]^2
       \left[ \overline{S}(q)-1 \right] \,,
\ee
where $L_L(x)$ are Laguerre polynomials and $L$ corresponds to the LL
index.

As in any MC calculation using the the Metropolis algorithm
\cite{MMC},  the expectation value of any [position dependent, e.g.\
$\rho(\rr)$] operator can be computed by averaging the local value of the
operator over a large number of electronic configurations generated
from the probability distribution $P \propto |\Psi_{1/m}^{\alpha}|^2$.
In a MC attempt, one eletron is moved to a new position
$\rr_{trial}=\rr_i+\Delta_i$, where $\Delta_i$ is a random vector in
some domain. If the probability ratio, $P(\rr_{trial})/P(\rr_i)$ is
larger than a random number uniformly distributed in the [0,1] range
then the move is accepted and we let let $\rr_{i+1}=\rr_{trial}$,
otherwise the move is rejected and $\rr_{i+1}=\rr_{i}$.  We adjust the
size of the domain over which $\Delta_i$'s vary so that about half of
the attempted moves are accepted.  Following standard practice, we
denote a MC step (MCS) a sequence of steps described above so that
every electron in the system has attempted a move (and about half
succeed).  After a MCS the system is in a state essentially
uncorrelated to the previous one and averages are computed for the
desired operators \cite{averages}
The results we report were obtained after discarding 100,000
``thermalization'' MCS's and using between $2 \times 10^6$ and $4
\times 10^7$ MCS's for averaging purposes on systems of 200--400
electrons.

\section{Monte Carlo results and discussion}
\label{sec:results}

By using MC methods we studied the possibility of a liquid crystal
state in the LLL for the leading candidate states at filling factors,
$\nu=1/3$, $1/5$ and $1/7$.  A trial wave function as in Eq.\
(\ref{eq:psi_nematic}) was considered and various properties were
analyzed as function of the anisotropic parameter $\alpha$.  Various
interaction potentials were considered for the computation of the
correlation energies [see Eqs.\ (\ref{thickness}, \ref{2other},
\ref{eq:ener}, \ref{eq:energy}, \ref{eq:energy-struc},
\ref{eq:energy-struc-L})], all have in common the fact that they
incorporate the effects of finite layer thickness into the quasi-2D
electronic system and are essentially identical to Coulomb's for large
distances. This choice is motivated by the wellknown fact that the
finite layer thickness of a real 2D system leads to a weakening and
eventual collapse of the FQHE \cite{he}.
Therefore, when the finite layer thickness (parameter $\lambda$) increases
as to become larger than the magnetic length, the short-range part of the
Coulomb interaction softens and as a result the isotropic FQHE liquid state
may become unstable with respect to another state of different nature
(a possible new candidate can be the liquid crystal state considered here,
and/or a Wigner crystal).

\begin{figure}[ht]
\begin{center}
\leavevmode
\includegraphics[width=3.3in]{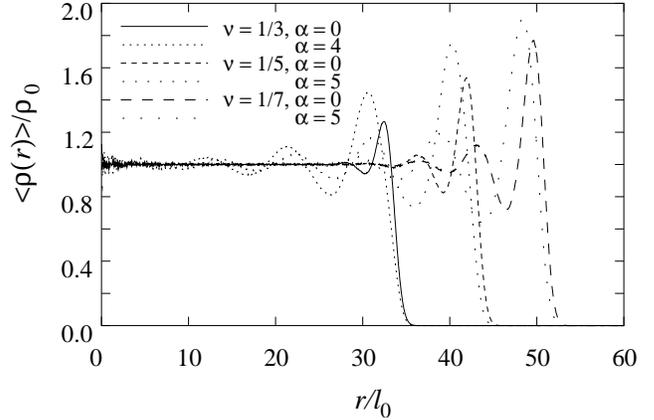}
\end{center}
\caption{\label{fig:rho}
Angle-averaged single-particle density, $\overline{\rho}(r)$, for $N =
196$ electrons and filling factors $\nu = 1/3$, 1/5 and 1/7.  We show
the results for the isotropic cases ($\alpha = 0$) and, for a large
$\alpha$ (the oscillations observed in this case are discussed in Sec.\
\ref{sec:cdw}). Here $r$ is the distance from the center of the disk.}
\end{figure}

In Fig. \ref{fig:rho} we show a plot of the angle-averaged
single-particle density $\overline{\rho}(r)$ for states of $N = 196$
electrons and filling factors of $\nu = 1/3$, 1/5 and 1/7.  The
existence, for $\alpha = 0$ of a large region around the center of the
disk ($r = 0$) with constant density is an indication that there is
bulk-like behavior \cite{mc_methods}.
Results for moderate values of $\alpha$ are similar to those for $\alpha=0$.
For larger $\alpha$ an apparent density fluctuation propagates from the edges
to the center making it very difficult to identify a ``bulk'' region.
The existence of this density fluctuation is discussed in detail in
Sec.\ \ref{sec:cdw}. We found that values of $\alpha$ acceptable for
the purposes of calculating bulk-like properties in reasonably sized
systems are as follows: $\alpha \lesssim  3$ for $\nu = 1/3$, $\alpha
\lesssim 4$ for both $\nu = 1/5$ and $1/7$ respectively.

\begin{figure}[ht]
\begin{center}
\leavevmode
\includegraphics[width=6.6in]{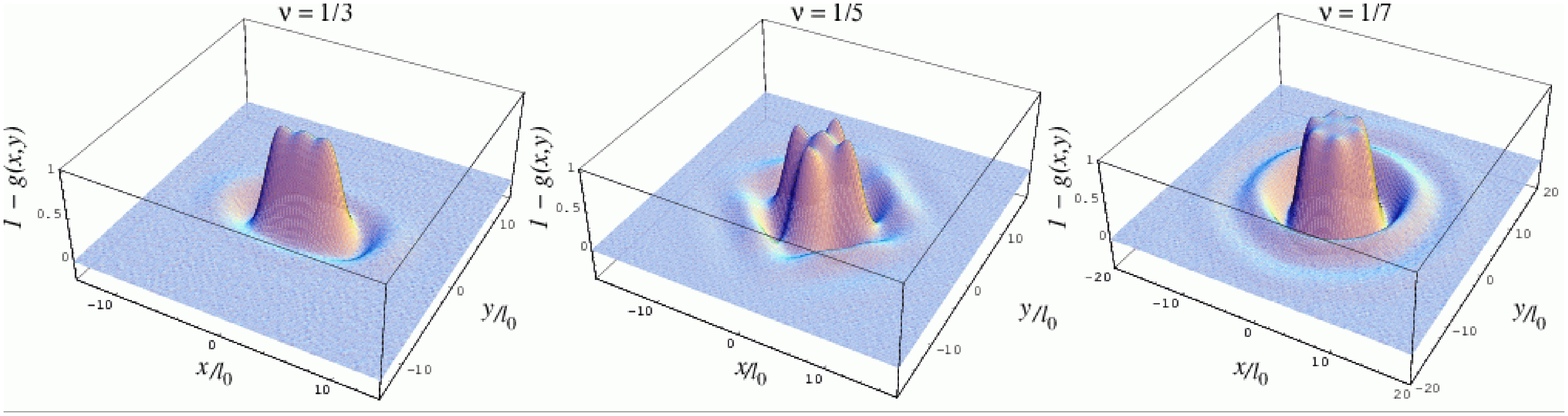}
\end{center}
\caption{\label{fig:gxy}
Pair correlation function $g(\rr)$ for $\nu = 1/3, \alpha = 2$ (left
panel), $\nu = 1/5, \alpha = 3$ (center panel), $\nu = 1/7, \alpha = 3$
(right panel). Note the discrete rotational symmetry of each state.}
\end{figure}
\begin{figure}[ht]
\begin{center}
\leavevmode
\includegraphics[width=6.6in]{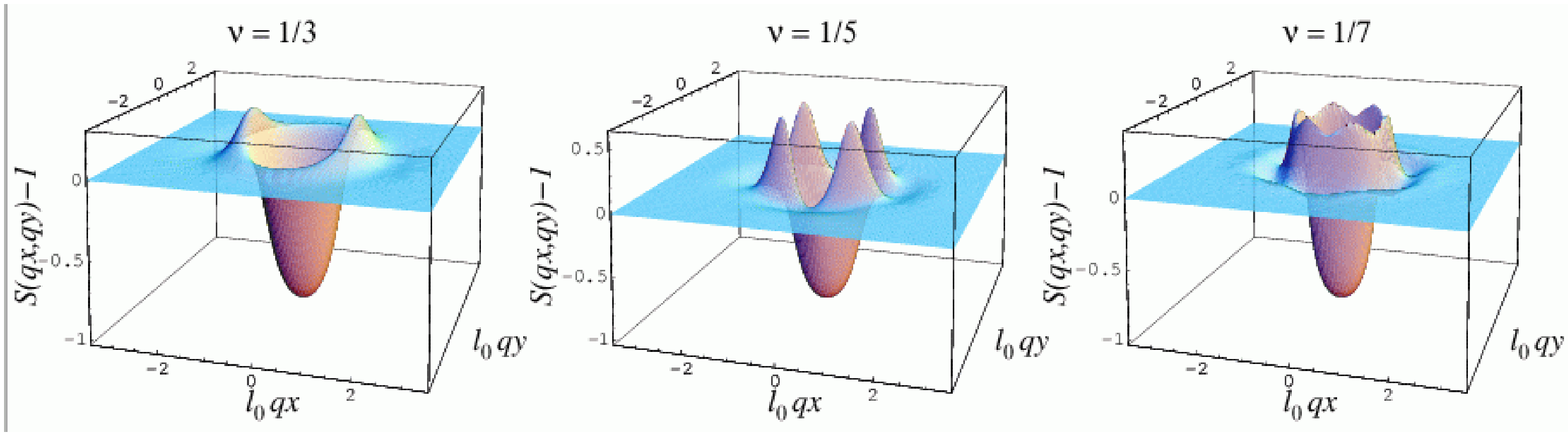}
\end{center}
\caption{\label{fig:sqxqy}
Static structure factor $S(\qq)$ for $\nu = 1/3, \alpha = 2$ (left
panel), $\nu = 1/5, \alpha = 3$ (center panel), $\nu = 1/7, \alpha = 3$
(right panel). Note the discrete rotational symmetry of each state.}
\end{figure}

In order to compare the energy of the isotropic Laughlin liquid state
with that of an anisotropic liquid crystal state, we first need an
accurate  computation of the pair distribution function
in terms of the parameter $\alpha$.  For the smallest $\alpha$'s,
a number of $N=196$ electrons was sufficient to give a very accurate pair
distribution function, whereas as many as 400 electrons were used when
$\alpha$-s became large as to induce sizeable oscillations in the
density.  Figure \ref{fig:gxy} shows results for the pair distribution
function, $g(\rr)$, for the: $\nu = 1/3$, $\alpha = 2$ nematic,
$\nu = 1/5$, $\alpha = 3$ tetratic, and $\nu = 1/7$, $\alpha = 3$ hexatic.
Each MC simulation involved $4 \times 10^7$ MCS's and ca.\ 400 electrons.
Figure \ref{fig:sqxqy} shows the corresponding static structure
factors $S(\qq)$ obtained from $g(\rr)$ using Eq.\ (\ref{eq:sq}).

\end{multicols}

\begin{figure}[ht]
\begin{center}
\leavevmode
\includegraphics[width=6.6in]{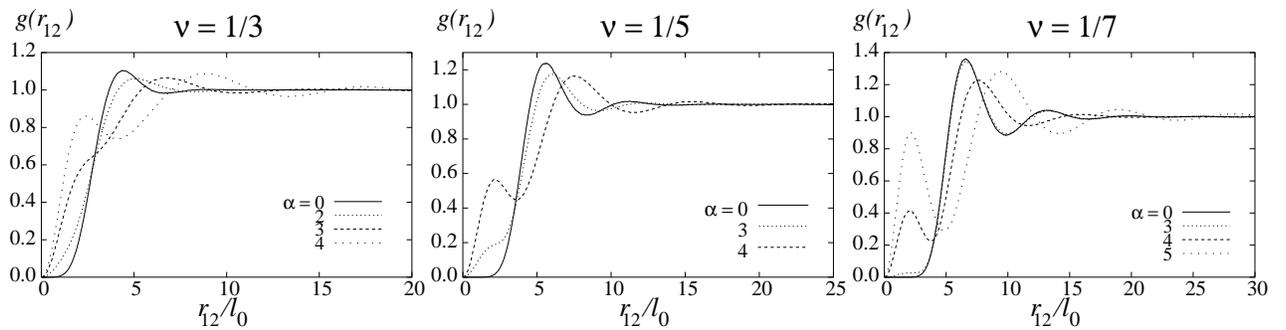}
\end{center}
\caption{\label{fig:gr}
Angle-averaged pair correlation function $\bar{g}(r_{12})$
for $\nu = 1/3$,  $\nu = 1/5$ and $\nu = 1/7$. }
\end{figure}

Since the angle-averaged $\bar{g}(r_{12})$ is sufficient for the
determination of the energy, we averaged it (at significant
savings in computer time) for various combinations of filling
factor $\nu$, and anisotropy parameter $\alpha$.  Figure \ref{fig:gr}
shows some of our results for 196--400 electrons.


\begin{multicols}{2}

At all filling factors that we considered, we noted that
$\overline{g}(r_{12})$ changes very little when parameter $\alpha$ is
small (e.g.\ 1).  Only for larger $\alpha$'s ($\gtrsim 2$) sizeable changes
take effect.  In view of this behavior, we anticipate that the energy
differences between the isotropic liquid state ($\alpha=0$) and the
anisotropic liquid crystal state with small anisotropy parameters
($\alpha=1$) will be quite small.  In fact, the calculation of energy
differences between these states and the isotropic state are comparable to
the estimated accuracy of our energy calculations. However, since the energy
differences for larger $\alpha$'s show a definite tendency in all cases, we
believe that the results are, significantly reliable
(since the statistical uncertainty on any MC calculation is systematic,
the energy differences may be even more accurate than the absolute
energies).

\begin{figure}
\begin{center}
\leavevmode
\includegraphics[width=6.6in]{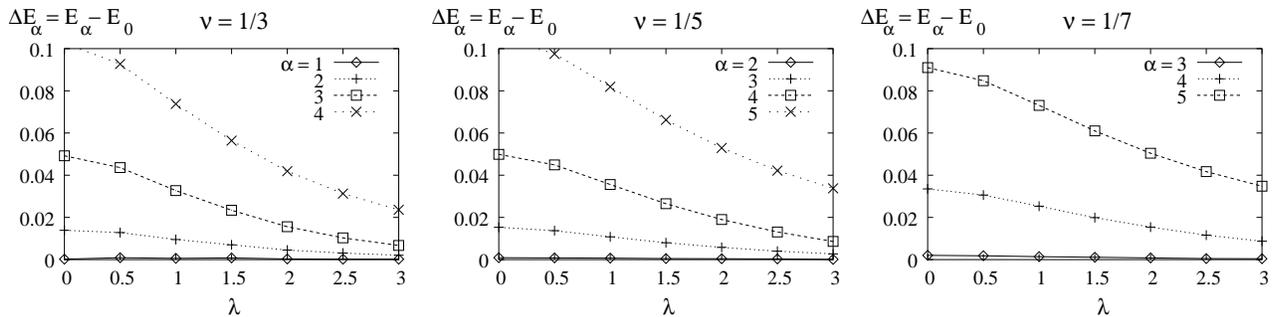}
\end{center}
\caption{\label{fig:DE13L0}
Energy difference between ansiotropic states and the isotropic state
($\alpha = 0$) $\Delta E_\alpha \equiv E_\alpha - E_0$ for filling
factors $\nu = 1/3$, 1/5 and 1/7 in the LLL.  These results correspond
to the interaction potential $v_1(r_{12})$ and are plotted as function of
the quasi-2D layer thickness $\lambda$.
}
\end{figure}

Tables \ref{table1}, \ref{table2} and \ref{table3} present the results
for the calculation of the LLL correlation energies obtained by means of
Eqs.\ (\ref{eq:energy}) or (\ref{eq:energy-struc}), using the
angle averaged pair correlation functions (or static structure
factors) for the three different forms of the interaction potential
for a variety of quasi-2D layer widths $\lambda$ [see Eqs.\
(\ref{thickness},\ref{2other})].  When $\lambda=0$ all interaction
potentials reduce to the Coulomb potential and in the case of the
$v_{ZDS}(r_{12})$ potential we note that $b=1.5/\lambda$. Results for
filling factors $\nu = 1/3$, 1/5 and 1/7 of the LLL (for the potential
$v_1(r_{12})$) are also presented in Fig.\ \ref{fig:DE13L0}.
The results suggest that, in the LLL, for all the interaction
potentials under consideration, a uniform liquid state is
energetically more favorable than the liquid crystal state.
For small values of $\alpha \in (0,\approx 2]$, the liquid crystal
states have an energy only slightly above the Laughlin liquid
states ($\alpha=0$), however for larger $\alpha$'s this difference
increases.

\end{multicols}
\begin{multicols}{2}

\begin{figure}
\begin{center}
\leavevmode
\includegraphics[width=3.3in]{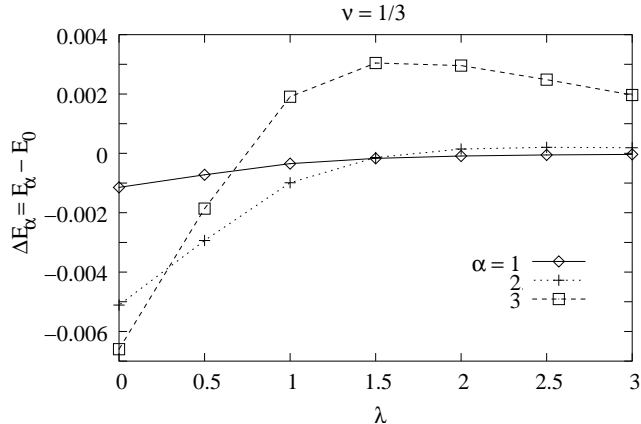}
\end{center}
\caption{\label{fig:DE13L2}
Energy difference between ansiotropic states and the isotropic state
($\alpha = 0$) $\Delta E_\alpha \equiv E_\alpha - E_0$ for filling
factor $\nu = 1/3$ in the second excited valence LL ($L = 2$).
These results correspond to the interaction potential form $v_1(r_{12})$
and are plotted as function of the quasi-2D layer thickness $\lambda$.
}
\end{figure}

Similar results are obtained in the first excited LL [$L = 1$ in Eq.\
(\ref{eq:energy-struc-L}), we omit the results for brevity].  For all
forms of the interaction potential considered here, the correlation energy for
anisotropic states is higher, once again leaving the Laughlin state
stable. However, it is interesting to note that for the second excited
LL ($L = 2$) the situation changes for the nematic states at $\nu =
1/3$ of the valence LL, where anisotropic states become energetically
favorable. 
Table \ref{table:LL2} shows the results for the energies, $E_\alpha$
and energy differences, $\Delta E_\alpha \equiv E_\alpha - E_0$ (also
shown in Fig.\ \ref{fig:DE13L2}) between anisotropic states ($\alpha
\neq 0$) and the isotropic state ($\alpha = 0$)
for filling factor $\nu = 1/3$ in the second excited LL ($L = 2$)
obtained from potential $v_1(r_{12})$ (the results are quite similar for
the other two forms of the potential).
These results are generally consistent to what we found in
the past using the hypernetted-chain (HNC) approximation
\cite{brsthird,brshalf}.

A conclusion can be derived from the above results: generally
speaking the isotropic states seem to be energetically favorable, with
the exception of the nematic state in the second excited LL.  The
explanation for this is simple:  in the LLL the electron packets
are simple gaussians, and it is clear that the best way to minimize
their Coulomb repulsion is by placing the vortices responsible for the CF
transformation \cite{heinonen,jain} precisely at the
location of the electron themselves ($\alpha = 0$).  In higher LL's,
the wavepackets take a more ``ring-like'' shape, and a finite $\alpha$
permits a more optimal distribution of charge for the nematic case
(but {\em not} for either the tetratic or hexatic).

\section{Underlying charge density wave in the anisotropic 2DOCP's}
\label{sec:cdw}

In view of the appearance of considerable density variations in our MC
simulations for larger values of $\alpha$ we investigated the possible
existence of an underlying CDW for the liquid crystalline states of Eq.\
(\ref{eq:psi_nematic}).  For this purpose it is useful to consider, once
again, the 2DOCP analog system.  Whereas considerable effort has been
dedicated (and a consequent vast knowledge has been achieved) in the
past to the treatment of the standard {\em isotropic} plasma (see
e.g.\ Refs.\ \onlinecite{caillol,debye2DOCP,hnc2d,sumrules}), little
has been pursued for a system with anisotropic interactions, e.g.\
quadrupolar terms. 

Consider the classical distribution function (note: in this section we
work in units of the magnetic length $l_0$):
\bea
\label{2docpANISO}
|\Psi_{1/m}|^2 &\propto& e^{-\beta V}  \ , \ {\rm where}\\
-\beta V &=& 2 \sum_{i<j}^{N} \left[
    \ln |z_i-z_j| + \sum_{\mu=1}^{m-1} |z_i-z_j-\alpha_\mu|
    \right] \nn \\
   && -\frac{1}{2}
             \sum_{k=1}^{N} |z_k|^2  \ , \nn
\eea
where, as before,
$\alpha_\mu = \alpha \, e^{i \theta_\mu}$,
$\theta_\mu = 2\pi(\mu\!-\!1)/(m\!-\!1)$, and $\mu \in \{1,2,\ldots,
(m-1)\}$. 
This potential energy corresponds to an ``electrostatic potential''
which is solution of a modified Poisson's equation
\be
\label{eq:poisson}
\nabla^2 [\beta \phi(\rr)] = - 4 \pi \Bigl[ \rho(\rr)
+ \sum_{\mu=1}^{m-1} \rho(\rr - \vec{\alpha}_\mu) \Bigr]
+ 4 \pi m \rho_0
\,,
\ee
where $\vec{\alpha}_\mu = \alpha (\cos \theta_\mu, \sin\theta_\mu)$,
\be
\label{eq:rhosource}
\rho(\rr) = \sum_{i=1}^{N} \delta(\rr - \rr_i) \,,
\ee
and $\rho_0 = 1/(2 \pi m)$  is a neutralizing density.

Consider now the potential $V$ generated by the addition of some charge
$\delta \rho(\rr)$.  This will cause a redistribution of the particles that
form the plasma, inducing a density change [see the discussion related
to the definition of the pair correlation function, Eq.\ (\ref{eq:gr})].
\be
\label{eq:rhoind}
\rho_{ind}(\rr) = \int d^2r' \, \rho_0 [g(\rr - \rr')-1] \,
 \delta \rho(\rr')\,.
\ee
The total charge, in reciprocal space, is therefore given by [see Eq.\
(\ref{eq:sq})]:
\be
\label{eq:rhototrecip}
\widetilde{\rho}_{tot} (\kk) = S(\kk) \, \widetilde{\delta \rho}(\kk) \,,
\ee
leading to a total potential:
\be
\label{eq:Vtot}
\beta \widetilde{\phi}(\kk) = \frac{4 \pi \, S(\kk)}{k^2}
    \left[ 1 + \sum_{\mu=1}^{m-1} e^{i \vec{\alpha}_\mu \cdot \kk} \right]
    \widetilde{\delta \rho}(\kk)
    \,.
\ee
This result neglects second order corrections in the distribution
functions and is, therefore, commonly referred to as the theory of
linear screening. 

It is now interesting to investigate whether this potential allows for
the formation of underlying CDW's in the 2DOCP.  Assuming small
variations from a uniform state, we allow for the particle density to
vary from point to point according to: 
\be
\label{eq:rhocdw}
\rho(\rr) = \rho_0 + \rho_1 \, \cos (\qq \cdot \rr) \,,
\ee
where $\qq$ is the wavevector of the CDW and $\rho_1 \ll \rho_0$.
The ${\cal O}[ \rho_1^2]$ ``excess energy'' \cite{caillol}
per unit area is given by:
\be
\label{eq:excessE}
\frac{\beta  u^{exc}}{\rho_1^2} =
\frac{1}{2} \, \frac{2 \pi \, S(\qq)}{q^2}
\left[ 1  + \sum_{\mu=1}^{m-1} e^{i \vec{\alpha}_\mu \cdot \qq} \right] \,.
\ee

\begin{figure}[ht]
\begin{center}
\leavevmode
\includegraphics[width=6.6in]{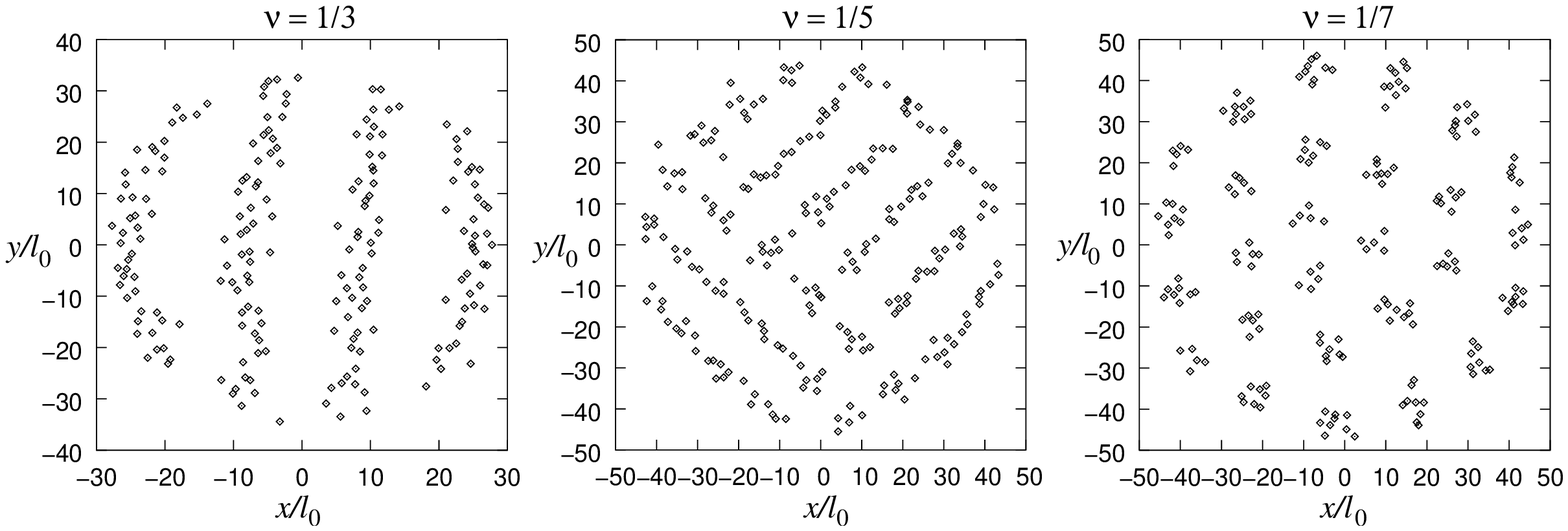}
\end{center}
\caption{\label{fig:cdw}
Typical electron configurations for a nematic ($\nu = 1/3$, $\alpha =
7$, left panel), tetratic ($\nu = 1/5$, $\alpha = 8$, center panel),
and nematic ($\nu = 1/7$, $\alpha = 10$, right panel). 
Note the formation of a CDW's with one, two and three different directors.
}
\end{figure}

It is evident that the charge neutrality sum rule ($S(\qq) \propto q^2$ for
$q\rightarrow 0$ \cite{caillol,SMA,sumrules}) guarantees the elimination
of the singularity at $q = 0$ leading to screening of the interaction.
More interesting, however, is the fact that the excess energy becomes
negative for a variety of wavevectors when $\alpha \neq 0$.  If we
write Eq.\ (\ref{eq:excessE}) explicitly for the various
states considered in this paper:
%
\end{multicols}
\bea
&&\text{nematic} {\; (\nu = 1/3):}\  \
    \frac{1}{2}\, \frac{2\pi \, S(\qq)}{q^2} 
	[1 + 2 \cos(\alpha q_x)] \,,  \nn \\
&&\text{tetratic} {\; (\nu = 1/5):}\  \
    \frac{1}{2} \, \frac{2\pi \,S(\qq)}{q^2}
        [1 + 2 \cos(\alpha q_x) + 2 \cos(\alpha q_y)] \,, \\
&&\text{hexatic} {\; (\nu = 1/7):}\  \
    \frac{1}{2} \, \frac{2\pi \,S(\qq)}{q^2} [1 + 2 \cos(\alpha q_x)
+ 2 \cos[\alpha (-\frac{1}{2}q_x + \frac{\sqrt{3}}{2} q_y)] +
+ 2 \cos[\alpha (-\frac{1}{2}q_x - \frac{\sqrt{3}}{2} q_y)] 
    ] \,,  \nn
\eea
We can see that the most important configurations (those that make the
potential $\beta V$ minimum and maximize their probability) correspond
to charge density waves with wavevectors in the neighborhood
of \cite{numden}
${\alpha \qq}/{\pi} \simeq \{(1,0), (-1,0)\}$ for the nematic,
${\alpha \qq}/{\pi} \simeq \{(1,1), (1,-1), (-1,1), (-1,-1) \}$ for
the tetratic, and ${\alpha \qq}/{\pi} \simeq 
\{  (\frac{4}{3}, 0), (-\frac{4}{3}, 0), (\frac{2}{3}, \frac{2}{\sqrt{3}}),
(\frac{2}{3}, -\frac{2}{\sqrt{3}}), (-\frac{2}{3}, \frac{2}{\sqrt{3}}),
(-\frac{2}{3}, -\frac{2}{\sqrt{3}}) \}$ for the hexatic.  
This should produce a unidirectional CDW (a layered system, or {\em
smectic}) underlying the nematic, 
with a characteristic wavelength $\lambda \simeq 2 \alpha$;
a square lattice tilted $45^\circ$ with lattice constant
$a \simeq \sqrt{2} \alpha$, and a triangular lattice with triangle side
$a = {\sqrt{3}} \alpha$. Figure \ref{fig:cdw} depicts typical
configurations during MC simulations with large $\alpha$'s.  The
characteristic CDW's have periods very close to those predicted above.

One should note that these underlying CDW's are extremely soft
and fluctuations will render them invisible in the thermodynamic and
ergodic limits.  In our simulations, however, their effects
are perceptible (see e.g.\ Fig.\ \ref{fig:rho}) for large values of
the anisotropy parameter $\alpha$ because of phase locking at the
boundaries. A detailed study of the fluctuations of these CDW's will
be published elsewhere \cite{cdwfl}.


\begin{multicols}{2}

\section{Conclusions}
\label{sec:conclusions}

In conclusion, we have investigated the possibility of liquid crystal
states in quasi-two-dimensional electron systems in strong magnetic
fields.  We considered translation invariant yet anisotropic states at
filling factors  $\nu = 1/3$, $1/5$ and $1/7$ of the  lowest ($L =
0$), first excited ($L = 1$) and second excited ($L = 2$) LL's.  We
found that the anisotropic states posess an underlying CDW along
directors with the same symmetry group of the proposed state but these
CDW's are ``washed-out'' by fluctuations.  We applied MC methods to
calculate the (angle-dependent) pair correlation function and static
structure factors for these states, which have permitted us to
calculate the correlation energies for a variety of reasonable
generalizations of the Coulomb potential that take into consideration
the finite width of the quasi-2D layer. For all states and potentials
under consideration the isotropic Laughlin state is found to be
energetically favorable in the lowest and first excited LL, whereas we
find an instability of the $\nu = 1/3$ nematic state in the second
excited LL.


\acknowledgements
We would like to thank A.T.\ Dorsey, A.H.\ MacDonald, E.\ Fradkin,
M.\ Fogler and J.\ MacCollough for useful discussions.
Acknowledgement is made to the University of Missouri Research Board
and to the Donors of the Petroleum Research Fund, administered by the
American Chemical Society, for support of this research.

One of the authors (OC) would like to acknowledge R.\ Wilkins and K.\
Kerby for their hospitality during summer 2003. Part of the work was
supported by NASA-CARR at Prairie View A\&M University.

\end{multicols}
\vspace{5cm}
\begin{multicols}{2}
\newpage

\begin{table}
\caption[]{ Correlation energy per particle in the LLL
            (in units of $e^2/\epsilon l_0$) for the liquid crystal
            (BRS) states at filling factor $\nu=1/3$ as a function
            of the anisotropy parameter $\alpha$ and quasi-2D layer width
            $\lambda$.  Three forms of the interaction potential were used.
            The three potentials reduce to the standard Coulomb potential
            for $\lambda=0$. }
\begin{center}
\begin{tabular}{|c|c|c|c|c|c|c|c|}
\hline
\multicolumn{8}{|c|}{ Interaction Potential: $v_{1}(r_{12})$  } \\
\hline
$\alpha$  &$\lambda=0.0$  &$\lambda=0.5$  &$\lambda=1.0$   &$\lambda=1.5$
          &$\lambda=2.0$  &$\lambda=2.5$   &$\lambda=3.0$
                                          \\ \hline
0    &-0.4100  &-0.3362  &-0.2776  &-0.2327  &-0.1973  &-0.1700 &-0.1485
   \\ \hline
1    &-0.4098  &-0.3353  &-0.2770  &-0.2319  &-0.1970  &-0.1698 &-0.1483
  \\ \hline
2    &-0.3961  &-0.3234  &-0.2681  &-0.2257  &-0.1928  &-0.1669 &-0.1464
   \\ \hline
3    &-0.3608  &-0.2926  &-0.2449  &-0.2093  &-0.1817  &-0.1597 &-0.1418
   \\ \hline
4    &-0.3074  &-0.2435  &-0.2038  &-0.1763  &-0.1554  &-0.1387  &-0.1249
   \\ \hline
\multicolumn{8}{|c|}{ Interaction Potential: $v_{2}(r_{12})$  } \\
\hline
$\alpha$  &$\lambda=0.0$  &$\lambda=0.5$  &$\lambda=1.0$   &$\lambda=1.5$
          &$\lambda=2.0$  &$\lambda=2.5$   &$\lambda=3.0$
                                          \\ \hline
0   &-0.4100  &-0.3286  &-0.2598  &-0.2107  &-0.1760  &-0.1507 &-0.1315
   \\ \hline
1    &-0.4098  &-0.3277  &-0.2593  &-0.2104  &-0.1758  &-0.1505  &-0.1314
  \\ \hline
2    &-0.3961  &-0.3162  &-0.2519  &-0.2058  &-0.1727  &-0.1483  &-0.1297
   \\ \hline
3    &-0.3608  &-0.2859  &-0.2324  &-0.1936  &-0.1650  &-0.1433  &-0.1264
   \\ \hline
4    &-0.3074  &-0.2370  &-0.1950  &-0.1661  &-0.1439  &-0.1265  &-0.1125
   \\ \hline
\multicolumn{8}{|c|}{ Interaction Potential: $v_{ZDS}(r_{12})$  } \\
\hline
$\alpha$  &$\lambda=0.0$  &$\lambda=0.5$  &$\lambda=1.0$   &$\lambda=1.5$
          &$\lambda=2.0$  &$\lambda=2.5$   &$\lambda=3.0$
                                          \\ \hline
0   &-0.4100  &-0.3279  &-0.2748  &-0.2381   &-0.2112  &-0.1904 &-0.1738
   \\ \hline
1    &-0.4098  &-0.3270  &-0.2741  &-0.2376  &-0.2107  &-0.1900  &-0.1735
  \\ \hline
2    &-0.3961  &-0.3160  &-0.2657  &-0.2310  &-0.2053  &-0.1854  &-0.1696
   \\ \hline
3    &-0.3608  &-0.2873  &-0.2439  &-0.2138  &-0.1914  &-0.1739  &-0.1597
   \\ \hline
4    &-0.3074  &-0.2408  &-0.2054  &-0.1813  &-0.1633  &-0.1491  &-0.1375
   \\ \hline
\end{tabular}
\end{center}
\label{table1}
\end{table}

\begin{table}[h]
\caption[]{ Correlation energy per particle in the LLL
            (in units of $e^2/\epsilon l_0$) for the liquid crystal
            (BRS) states at filling factor $\nu=1/5$ as a function
            of the anisotropy parameter $\alpha$ and quasi-2D layer width
            $\lambda$.  Three forms of the interaction potential were used.
            The three potentials reduce to the standard Coulomb potential
            for $\lambda=0$. }
\begin{center}
\begin{tabular}{|c|c|c|c|c|c|c|c|}
\hline
\multicolumn{8}{|c|}{ Interaction Potential: $v_{1}(r_{12})$  } \\
\hline
$\alpha$  &$\lambda=0.0$  &$\lambda=0.5$  &$\lambda=1.0$   &$\lambda=1.5$
          &$\lambda=2.0$  &$\lambda=2.5$   &$\lambda=3.0$
                                          \\ \hline
0   &-0.3274  &-0.2811   &-0.2420  &-0.2094  &-0.1825  &-0.1603 &-0.1419
   \\ \hline
1    &-0.3273  &-0.2810  &-0.2419  &-0.2094  &-0.1825  &-0.1603  &-0.1419
  \\ \hline
2    &-0.3265  &-0.2803  &-0.2413  &-0.2089  &-0.1821  &-0.1600  &-0.1418
   \\ \hline
3    &-0.3121  &-0.2674  &-0.2312  &-0.2014  &-0.1767  &-0.1563 &-0.1392
   \\ \hline
4    &-0.2775  &-0.2362  &-0.2064  &-0.1829  &-0.1635  &-0.1472  &-0.1333
   \\ \hline
5    &-0.2216  &-0.1836  &-0.1601  &-0.1432  &-0.1296  &-0.1181  &-0.1081
   \\ \hline
\multicolumn{8}{|c|}{ Interaction Potential: $v_{2}(r_{12})$  } \\
\hline
$\alpha$  &$\lambda=0.0$  &$\lambda=0.5$  &$\lambda=1.0$   &$\lambda=1.5$
          &$\lambda=2.0$  &$\lambda=2.5$   &$\lambda=3.0$
                                          \\ \hline
0   &-0.3274  &-0.2743  &-0.2365  &-0.2086  &-0.1873  &-0.1704  &-0.1566
   \\ \hline
1    &-0.3273  &-0.2743  &-0.2365  &-0.2086  &-0.1873  &-0.1704  &-0.1566
  \\ \hline
2    &-0.3265  &-0.2767  &-0.2303  &-0.1928  &-0.1641  &-0.1422  &-0.1251
   \\ \hline
3    &-0.3121  &-0.2639  &-0.2215  &-0.1870  &-0.1603  &-0.1396  &-0.1233
   \\ \hline
4    &-0.2775  &-0.2329  &-0.1997  &-0.1730  &-0.1513  &-0.1338  &-0.1196
   \\ \hline
5    &-0.2216  &-0.1801  &-0.1561  &-0.1378  &-0.1222  &-0.1088  &-0.0975
   \\ \hline
\multicolumn{8}{|c|}{ Interaction Potential: $v_{ZDS}(r_{12})$  } \\
\hline
$\alpha$  &$\lambda=0.0$  &$\lambda=0.5$  &$\lambda=1.0$   &$\lambda=1.5$
          &$\lambda=2.0$  &$\lambda=2.5$   &$\lambda=3.0$
                                          \\ \hline
0   &-0.3274   &-0.2743  &-0.2365  &-0.2086  &-0.1873  &-0.1704  &-0.1566
   \\ \hline
1    &-0.3273  &-0.2743  &-0.2365  &-0.2086  &-0.1873  &-0.1704  &-0.1566
  \\ \hline
2    &-0.3265  &-0.2736  &-0.2359  &-0.2082  &-0.1869  &-0.1701 &-0.1563
   \\ \hline
3    &-0.3121  &-0.2615  &-0.2265  &-0.2006  &-0.1807  &-0.1648 &-0.1519
   \\ \hline
4    &-0.2775  &-0.2324  &-0.2037  &-0.1825  &-0.1659  &-0.1525  &-0.1414
   \\ \hline
5    &-0.2216  &-0.1819  &-0.1598  &-0.1439  &-0.1313  &-0.1211  &-0.1124
   \\ \hline
\end{tabular}
\end{center}
\label{table2}
\end{table}

\begin{table}[h]
\caption[]{ Correlation energy per particle in the LLL
            (in units of $e^2/\epsilon l_0$) for the liquid crystal
            (BRS) states at filling factor $\nu=1/7$ as a function
            of the anisotropy parameter $\alpha$ and quasi-2D layer width
            $\lambda$.  Three forms of the interaction potential were used.
            The three potentials reduce to the standard Coulomb potential
            for $\lambda=0$. }
\begin{center}
\begin{tabular}{|c|c|c|c|c|c|c|c|}
\hline
\multicolumn{8}{|c|}{ Interaction Potential: $v_{1}(r_{12})$  } \\
\hline
$\alpha$  &$\lambda=0.0$  &$\lambda=0.5$  &$\lambda=1.0$   &$\lambda=1.5$
          &$\lambda=2.0$  &$\lambda=2.5$   &$\lambda=3.0$
                                          \\ \hline
0   &-0.2827  &-0.2491  &-0.2198  &-0.1944  &-0.1727  &-0.1541 &-0.1383
   \\ \hline
1    &-0.2827  &-0.2491  &-0.2198  &-0.1944  &-0.1727  &-0.1541  &-0.1383
  \\ \hline
2    &-0.2826  &-0.2491  &-0.2198  &-0.1944  &-0.1727  &-0.1541  &-0.1383
   \\ \hline
3    &-0.2807  &-0.2473  &-0.2184  &-0.1933  &-0.1719  &-0.1536  &-0.1379
   \\ \hline
4    &-0.2492  &-0.2185  &-0.1945  &-0.1745  &-0.1573  &-0.1425  &-0.1296
   \\ \hline
5    &-0.1917  &-0.1643  &-0.1467  &-0.1334  &-0.1223  &-0.1124  &-0.1035
   \\ \hline
\multicolumn{8}{|c|}{ Interaction Potential: $v_{2}(r_{12})$  } \\
\hline
$\alpha$  &$\lambda=0.0$  &$\lambda=0.5$  &$\lambda=1.0$   &$\lambda=1.5$
          &$\lambda=2.0$  &$\lambda=2.5$   &$\lambda=3.0$
                                          \\ \hline
0   &-0.2827  &-0.2470   &-0.2123  &-0.1821  &-0.1576  &-0.1382  &-0.1227
   \\ \hline
1    &-0.2827  &-0.2470  &-0.2124  &-0.1822  &-0.1577  &-0.1383  &-0.1227
  \\ \hline
2    &-0.2826  &-0.2470  &-0.2123  &-0.1821  &-0.1576  &-0.1382  &-0.1227
   \\ \hline
3    &-0.2807  &-0.2452  &-0.2110  &-0.1813  &-0.1571  &-0.1378  &-0.1225
   \\ \hline
4    &-0.2492  &-0.2164  &-0.1894  &-0.1659  &-0.1460  &-0.1297  &-0.1162
   \\ \hline
5    &-0.1917  &-0.1621  &-0.1443  &-0.1295  &-0.1160  &-0.1040  &-0.0936
   \\ \hline
\multicolumn{8}{|c|}{ Interaction Potential: $v_{ZDS}(r_{12})$  } \\
\hline
$\alpha$  &$\lambda=0.0$  &$\lambda=0.5$  &$\lambda=1.0$   &$\lambda=1.5$
          &$\lambda=2.0$  &$\lambda=2.5$   &$\lambda=3.0$
                                          \\ \hline
0   &-0.2827   &-0.2436  &-0.2141  &-0.1914  &-0.1735  &-0.1591 &-0.1471
   \\ \hline
1    &-0.2827  &-0.2437  &-0.2141  &-0.1915  &-0.1736  &-0.1591  &-0.1472
  \\ \hline
2    &-0.2826  &-0.2436  &-0.2141  &-0.1914  &-0.1735  &-0.1591 &-0.1471
   \\ \hline
3    &-0.2807  &-0.2419  &-0.2128  &-0.1904  &-0.1727  &-0.1584 &-0.1465
   \\ \hline
4    &-0.2492  &-0.2147  &-0.1906  &-0.1720  &-0.1572  &-0.1450  &-0.1348
   \\ \hline
5    &-0.1917  &-0.1627  &-0.1454  &-0.1323  &-0.1216  &-0.1126  &-0.1050
   \\ \hline
\end{tabular}
\end{center}
\label{table3}
\end{table}

\begin{table}[h]
\caption[]{ Correlation energy per particle in the second excited LL,
    $L = 2$, (in units of $e^2/\epsilon l_0$) for the liquid crystal
        (BRS) states at filling factor $\nu=1/3$ as a function
        of the anisotropy parameter $\alpha$ and quasi-2D layer width
        $\lambda$.  The form $v_1(r_{12})$ for the interaction
	potential was used.  } 
\begin{center}
\begin{tabular}{|c|c|c|c|c|c|c|c|}
\hline
\multicolumn{8}{|c|}{ Interaction Potential: $v_{1}(r_{12})$  } \\
\hline
$\alpha$  &$\lambda=0.0$  &$\lambda=0.5$  &$\lambda=1.0$   &$\lambda=1.5$
          &$\lambda=2.0$  &$\lambda=2.5$   &$\lambda=3.0$
                                          \\ \hline
0  &  -0.2642  &  -0.2139  &  -0.1872  &  -0.1662  &  -0.1485  &
-0.1335  &  -0.1207  \\ \hline
1  &  -0.2653  &  -0.2146  &  -0.1875  &  -0.1663  &  -0.1486  &
-0.1335  &  -0.1208  \\ \hline
2  &  -0.2693  &  -0.2169  &  -0.1881  &  -0.1663  &  -0.1483  &
-0.1333  &  -0.1206  \\ \hline
3  &  -0.2708  &  -0.2158  &  -0.1852  &  -0.1631  &  -0.1455  &
-0.1310  &  -0.1188   \\
\hline
\end{tabular}
\end{center}
\label{table:LL2}
\end{table}


\newpage

\end{multicols}

\end{multicols}
\end{document}